\documentclass[12pt, margin = 1in]{article}
\PassOptionsToPackage{table}{xcolor}
\usepackage[utf8]{inputenc}
\usepackage[margin = 1in]{geometry}
\usepackage{times}
\usepackage{amsmath}
\usepackage{amsfonts}
\usepackage{setspace}
\usepackage{newtxtext,newtxmath}
\usepackage{graphicx}
\usepackage{rotating}
\usepackage{authblk}
\usepackage{color}
\usepackage[round]{natbib}
\usepackage{tikz}
\usepackage{bm}
\usepackage{bigints}
\usepackage{multirow}
\usepackage{float}
\usepackage{longtable}
\usepackage{lscape}
\usepackage{setspace}   
\usepackage{mathtools, bbold}
\usepackage{subcaption}
\usepackage{makecell}
\usepackage[table]{xcolor} 
\usepackage[thinlines]{easytable}
\usepackage{diagbox}
\usepackage{booktabs}
\usepackage{dsfont}
\usepackage[OT1]{fontenc}
\usepackage[colorlinks,citecolor=blue,urlcolor=blue]{hyperref}
\usepackage[inline]{trackchanges}
\usetikzlibrary{shapes,decorations,arrows,calc,arrows.meta,fit,arrows,shapes.arrows,
shapes.geometric,shapes.multipart,decorations.pathmorphing,positioning,shapes.swigs}
\tikzset{
    -Latex,auto,node distance =1 cm and 1 cm,semithick,
    state/.style ={ellipse, draw, minimum width = 0.7 cm},
    point/.style = {circle, draw, inner sep=0.04cm,fill,node contents={}},
    bidirected/.style={Latex-Latex,dashed},
    el/.style = {inner sep=2pt, align=left, sloped}
}

\newcommand{\E}{\mathsf{E}}
\newcommand{\Esp}{\mathsf{E}}
\renewcommand{\d}{\mathsf{d}}
\renewcommand{\P}{\mathsf{P}}
\newcommand{\V}{\mathsf{Var}}
\newcommand{\cov}{\mathsf{cov}}
\newcommand{\ecov}{\mathsf{c\hat{o}v}}
\newcommand{\mathbold}{\bm}
\newcommand\independent{\protect\mathpalette{\protect\independenT}{\perp}}

\def\independenT#1#2{\mathrel{\rlap{$#1#2$}\mkern2mu{#1#2}}}

\newtheorem{remark}{Remark}

\DeclareUnicodeCharacter{2061}{}

  \title{Integration of aggregated data in causally interpretable meta-analysis by inverse weighting}
\author[1]{Tat Thang Vo \footnote{\footnotesize \textbf{Correspondence:} \\Tat Thang Vo, Research Group EPIDERME, Faculty of Medicine, University Paris Est Creteil, France.\\Email: tat-thang.vo@u-pec.fr}}
\author[1]{Tran Trong Khoi Le}
\author[1]{Sivem Afach}
\author[2]{Stijn Vansteelandt}
\affil[1]{Research group EPIDERME, Faculty of Medicine, University Paris Est Creteil, France}
\affil[2]{Department of Applied Mathematics, Computer Science and Statistics, Ghent University, Belgium}

\begin{document}
\maketitle
\setlength{\baselineskip}{12pt}


\begin{abstract}
Obtaining causally interpretable meta-analysis results is challenging when there are differences in the distribution of effect modifiers between eligible trials. To overcome this, recent work on transportability methods has considered standardizing results of individual studies over the case-mix of a target population, prior to pooling them as in a classical random-effect meta-analysis. One practical challenge, however, is that case-mix standardization often requires individual participant data (IPD) on outcome, treatments and case-mix characteristics to be fully accessible in every eligible study, along with IPD case-mix characteristics for a random sample from the target population. In this paper, we aim to develop novel strategies to integrate aggregated-level data from eligible trials with non-accessible IPD into a causal meta-analysis, by extending moment-based methods frequently used for population-adjusted indirect comparison in health technology assessment. Since valid inference for these moment-based methods by M-estimation theory requires additional aggregated data that are often unavailable in practice, computational methods to address this concern are also developed. We assess the finite-sample performance of the proposed approaches by simulated data, and then apply these on real-world clinical data to investigate the effectiveness of risankizumab versus ustekinumab among patients with moderate to severe psoriasis.\\
\end{abstract}

\noindent%

\singlespacing
\section{Introduction}
\label{sec:sec1}
Meta-analysis is a quantitative procedure of assessing and combining data from multiple studies \citep{chandler2019cochrane}. 
Conventional meta-analysis methods, however, do not provide findings with a clear causal interpretation when the distribution of effect modifiers differs between eligible trials \citep{dahabreh2020toward, truong2023population,vo2019novel}.
To remedy this, a novel framework called \textit{causally interpretable meta-analysis} (or \textit{causal meta-analysis} for short) has recently been proposed \citep{dahabreh2020toward,vo2019novel,vo2021assessing,sobel2017causal}. This framework considers standardizing results of eligible studies over the case-mix of a target population, before pooling them as in a classical random-effect meta-analysis. The obtained summary findings do not suffer from case-mix heterogeneity between studies, and quantify the treatment effect for a well-defined target population. 

One practical challenge of causal meta-analysis, however, is that it requires the individual participant data (IPD) to be fully accessible in every eligible study and in the target population. In practice, data sharing is often subject to administrative and privacy concerns \citep{ross2016clinical}. To the best of our knowledge, little work has been done to formally address the complexity due to limited access to IPD in causal meta-analysis. Quite recently, \citet{rott2024causally} has proposed a new method to simulate covariates for individuals of studies without IPD accessible, when certain summary statistics (or aggregated data) are available from these non-IPD trials. This method, however, requires a strong parametric assumption that the covariance of covariates (COC) is either the same in all eligible studies and in the target population, or the COC in the target population is a mixture of the COCs in eligible studies with IPD. Other simulation-based methods have also been proposed by \citet{remiro2022parametric} and \citet{phillippo2020multilevel}, under a similar assumption. Alternatively, in health technology assessment (HTA), moment-based methods have been used to transport treatment effects of one randomized trial with IPD to a target population with aggregated data, without having to impose homogeneity on the covariance matrix of the covariates in different trial populations \citep{signorovitch2012matching, truong2023population}. The extension of these methods to causal meta-analysis has not yet been investigated. Besides, establishing asymptotic inference for these so-called population-adjusted methods is quite challenging \citep{cheng2020statistical}. Common practice includes randomly drawing bootstrap samples from the trial's IPD, while considering the aggregated data of the target population as fixed. This overlooks the uncertainty in these aggregated data, and can lead to type I error inflation or under-coverage issues in the resulting confidence interval of the treatment effect estimate in the target population.

In this paper, we propose novel methods to integrate aggregated-level data from eligible trials with non-accessible IPD into a causal meta-analysis, with the aim to infer the summary treatment effect for each trial population in the analysis. To achieve this, we first revisit the moment-based methods for transportation of treatment effects across populations proposed in HTA, and then develop extensions of these methods to causal meta-analysis. Unlike previous methods, we do not require homogeneity of the covariance matrix of the covariates in different eligible trials to obtain consistent effect estimates. We also show that valid inference for these moment-based methods by M-estimation theory requires additional aggregated data that are often unavailable in practice. Methods to address this challenge are thus developed. We illustrate the proposed approaches on numerically simulated data, and on real-world data of a meta-analysis investigating the effectiveness of risankizumab versus ustekinumab among patients with moderate to severe psoriasis.

\section{Transportability of causal effects in meta-analysis}
\label{sec:sec2}
\subsection{Causal estimand, identification and estimation: a brief overview}
Consider a meta-analysis of  $q$  randomized trials to evaluate the comparative effectiveness of two treatments  ($X=0,1$)  on a dichotomous outcome  $(Y=0,1)$. Let $S$ be an indicator of the study from which a given patient originates, which takes values from 1 to $q$, and $\bm L = (L_1,\ldots, L_{p-1})$ be a set of $p-1$ baseline covariates that are collected across studies. Assume for now that the IPD is accessible for all eligible studies. In that case, we observe $O_i = (\bm L_i, S_i, X_i, Y_i) $ for each individual $i$ in the dataset. These observations $(O_1, \ldots, O_n)$ are independent and identically distributed realizations of $O = (\bm L, S, X, Y)$. 

Denote $Y \left( x,k \right)$ the counterfactual outcome that would be observed if a patient participated in trial $k$ and were assigned to treatment  $x$. Our focus is on identifying and estimating the risk difference:
\[ \theta(j,k) := \P\{ Y \left(x=1, {k} \right) =1 \mid S=j \} - \P \{ Y \left(x= 0, k \right) =1 \mid S=j \} \]
which expresses the treatment effect when all individuals from the target population of trial $S=j$ were given the (version of) treatment versus control used in trial $k$. While we here focus on risk differences, extension to other effect measures such as relative risks or odds ratios is relatively straightforward. To identify $\theta( j,k)$, the following assumptions are made for $x = 0,1$ and $j,k = 1, \ldots, q$:
\begin{enumerate}
\item [(i)] \textit{Transportability}, i.e., $Y \left( x,{k} \right) \independent S\mid \bm L$, which requires that $\bm L$ contains all outcome prognostic factors that are differentially distributed across populations. 
\item [(ii)] \textit{Positivity}, i.e. $\P\{0<\P \left( S=j \mid  \bm L \right) <1\} = 1$, which requires that trials are sufficiently similar in terms of target populations, so that there is an adequate overlap in the distribution of $\bm L$ across studies.
\item [(iii)] \textit{Consistency}, i.e. $Y \left( x,k \right) =Y$ when $S=k$ and $X=x$, which states that $Y (x,k)$ agrees with the observed outcome $Y$ for all individuals in study $k$ receiving treatment $x$. 
\item [(iv)] \textit{Ignorability within each trial}, i.e. $Y \left( x, {k} \right) \independent X \mid S$, which holds when the treatment is randomized within each study.
\end{enumerate}
A more detailed discussion of the above causal assumptions can be found in some recent works \citep{vo2019novel,sobel2017causal}. Under these assumptions, $\P\{ Y \left( x,{k} \right)=1\mid S=j\}$ can be linked to the observed data as:
\begin{equation}\label{eq2}
\P \left\{ Y \left( x,{k} \right) =1 \mid  S=j \right\} = \int _{}^{}\P \left( Y=1 \mid  X=x,\bm L=\bm l,S=k \right)  \d\P \left(\bm l \mid  S=j \right),
\end{equation}
This identity suggests that the variability across $\theta(j,k)$ can be explained by two factors, namely (i) differences across studies in the distribution of patient characteristics $\P(l|S=j)$ (also called \textit{case-mix heterogeneity}), and (ii) differences across studies in the outcome generating mechanism $\P(Y=1|X, L,S)$ (also called \textit{beyond case-mix heterogeneity}). 
\begin{remark}
In the transportability literature, the treatment and control versions are often implicitly assumed to be the same across trials, in the sense that $Y(x,k) = Y(x,j) = Y(x)$. In that case, $\theta(j,k) = \theta(j,j)$ will express the average treatment effect in the target population $j$, which can be identified by equation (\ref{eq2}) under a weaker transportability assumption, i.e. $Y(x) \independent S\mid \bm L$ \citep{degtiar2023review}. In practice, it is however more realistic to allow $Y(x,k) \ne Y(x,j)$, due to which rescaling results of trial $k$ over the case-mix of population $j$ can only reveal the effect of a version of treatment sufficiently similar to (but not neccessarily the same as) the treatment version evaluated in trial $j$. This so-called beyond case-mix heterogeneity is common in meta-analysis \citep{vo2019novel,vo2021assessing}. For instance, trials might have different non-compliance rates, or assess pharmacological treatments that are different in dose, route of administration and so forth.
\end{remark}
To estimate $\P\{Y(x,k)=1\mid S=j\}$, one simple approach is to postulate and fit a parametric outcome model for study $k$, then using such model to make outcome predictions under treatment $x$ for participants in study $j$, and averaging these predicted values (the \textit{outcome regression} or \textit{g-computation} approach).  Alternatively, one can consider an \textit{inverse weighting} approach based on a propensity score model for study membership, that is:
\[ \P \left\{ Y \left( x,{k} \right) =1 \mid  S=j \right\} =\frac{1}{\P \left( S=j \right) }\E \left\{ I \left( S=k \right)YI \left( X=x \right)\frac{\P \left(S=j \mid  \bm L \right) }{\P \left(S=k \mid \bm L \right) }\frac{1}{\P \left(X=x \mid S= k \right) } \right\}\]
Unlike g-computation, inverse weighting avoids assumptions on the degree of effect heterogeneity between populations, thereby reducing the potential risk of extrapolation when participants in different studies have very different case mix \citep{vo2023cautionary}. Other estimation approaches have also been proposed in the recent literature, which are more efficient and robust by allowing the use of flexible, data-adaptive methods to estimate the nuisance outcome and propensity score models \citep{vo2021assessing,dahabreh2023efficient}.

\subsection{Transportability with aggregated data}
We now extend the above framework to a more realistic setting where the IPD are not available for the first $z$ trials, i.e. $S=1,\ldots, z$ in the meta-analysis. The data structure is now $O = \big\{\bm LI(S>z), S, XI(S>z), YI(S>z)\big\}$. On many occasions, multiple summary statistics are nonetheless retrievable from the published report of trial $j=1,\ldots,z$. For $x=0,1$, these include:
\begin{enumerate}
\item[(i)] The first and second moments of $\bm L$ in each treatment group, i.e. $\hat \E (\bm L\mid X=x,S=j)$ and $\hat \E(\bm L^2\mid X=x,S=j)$, where $\bm L^2 = (L_1^2, \ldots , L_{p-1}^2)$.
\item[(ii)] The outcome mean in each treatment group, i.e. $\hat \E(Y\mid X=x,S=j)$.
\item[(iii)] The probability $r_{xj}$ of receiving treatment $x$, i.e. $r_{xj}=\P(X=x\mid S=j)$. 
\item[(iv)] The total number of patients in each treatment arm.
\item [(v)] The treatment effect estimate $\hat\theta(j,j)$, i.e. a consistent estimate of $\theta(j,j)$.
\end{enumerate}

Moreover, assume that we can correctly specify a parametric model for the propensity score $\P(S=s|\bm L)$, $s = 1, \ldots, q$, in the sense that:
\begin{equation}\label{eqw}
\frac{\P(S=j\mid \bm L)}{\P(S=k\mid \bm L)}= m(\bm L, \bm \beta_{jk})\quad \mathrm{for}\quad j,k = 1,\ldots, q
\end{equation}
and some finite-dimension parameter vector $\bm\beta_{jk}$. Below, we consider the following simple model: 
\[m(\bm L, \bm\beta_{jk})=\exp(\bm \beta_{jk}^\top \bm L) =\exp\big\{\beta_{jk}^{(0)} + \beta_{jk}^{(1)}L_1 + \ldots + \beta_{jk}^{(p-1)}L_{p-1}\big\}\] 
for educational purposes, but the proposed method is easily extended to more general settings (i.e. remark 2). 

When the IPD are available for both trials $j$ and $k$, standard approaches such as maximum likelihood can be used to estimate $\bm \beta_{jk}$. However,  these approaches are not applicable when the IPD are only accessible for either $S=j$ or $S=k$. Alternatively, note that:  
\begin{equation}\label{eq7.5}
\E \left\{ I(S=k)~\bm \phi (\bm L)~\frac{\P(S=j\mid \bm L)}{\P(S=k\mid \bm L)}\right\}=\P(S=j)~\E\{\bm \phi (\bm L)\mid S=j\} 
\end{equation}
for any arbitrary $p$-dimension vector function $\bm \phi(\bm L)$ of $\bm L$. This moment equation is particularly useful because solving its sample analog can provide a valid estimate $\hat{\bm\beta}_{jk}$ for $\bm\beta_{jk}$, and this only requires information on the sample mean of $ \bm \phi(\bm L)$ in trial $j$.
In particular, the matching-adjusted indirect comparison (MAIC) technique commonly used in HTA considers $ \bm \phi(\bm L) = \bm L^\top$.  Alternatively, one can also consider $\bm \phi(\bm L) = \begin{pmatrix} \bm L^\top & \bm L^{2,\top}\end{pmatrix}^\top$. This choice of $\bm\phi(\bm L)$ results in $2p$ estimating equations for a $p$-dimension parameter $\bm\beta_{jk}$. To obtain a unique estimate $\hat{\bm\beta}_{jk}$ for $\bm\beta_{jk}$, compute $\hat{\bm\beta}_{jk} = \mathrm{argmin}_{\bm \beta_{jk}} f(\bm O,\bm \beta_{jk})$, where:
\begin{align*} 
 f(\bm O,\bm \beta_{jk})=\bigg\Vert \sum_{i=1}^n I(S_i=k)~\bm \phi (\bm L_i)~m(\bm L_i, \bm\beta_{jk}) - \hat \E\{\bm \phi (\bm L)\mid S=j\} \bigg\Vert^2_2
\end{align*}
and $\Vert \cdot \Vert_2$ denotes the Euclidean norm. By construction, $\hat{\bm\beta}_{jk} \in \big\{\bm \beta_{jk}: \nabla f(\bm O,\bm \beta_{jk}) = \bm 0;~ -\nabla^2 f(\bm O,\bm \beta_{jk}) \succ 0 \big \}$, where $A \succ 0$ means $A$ is a positive-definite matrix, and $\nabla$ and $\nabla^2$ denote the first and second derivatives of a function with respect to the parameter vector $\bm\beta_{jk}$, respectively. The elements of this set are fully observed as computing the gradient and Hessian matrix of $f(\bm O,\bm \beta_{jk})$ only requires the IPD from trial $k$. 
\begin{remark}
Notice that the proposed moment-based approach enables some flexibility in specifying the linear predictor $m(\bm L, {\bm\beta}_{jk})$ of the propensity score model (\ref{eqw}). For instance, when: \[m(\bm L, {\bm\beta}_{jk}) = \beta_{jk}^{(0)} + \beta_{jk}^{(1)}L_1 + \beta_{jk}^{(2)}L_2 + \beta_{jk}^{(3)}L_1L_2,\] one may consider the vector function $\bm \phi(\bm L) = \begin{pmatrix} 1 & L_1 & L_2 & L_1^2 & L_2^2 \end{pmatrix}^\top$ to ensure that the number of estimating equations at least equals (or is larger than) the dimension of $\bm\beta_{jk}$. In general, since the first and second moment of $\bm L$ are available in each study $S=1,\ldots,z$, one can postulate and fit a parametric weight model with at most $2p-1$ unknown parameters, i.e. $|\bm\beta_{jk}|\le 2p-1$.
\end{remark}
Once model (\ref{eqw}) is estimated, the inverse weighting approach can be used to standardize results of trial $k$ with IPD over the case-mix of the target population of trial $j$ without IPD, since this approach only requires the individual-level data from trial $k$ to estimate $\E\{Y(x,k)\mid S=j\}$. 

\subsection{Statistical inference}
In this section, we derive the asymptotic distribution of the estimates $\hat\theta(j,k)$ obtained from the above strategy. To achieve this, denote $\bm \theta_k=\begin{pmatrix} \bm \beta_k^\top & \bm \delta_k^\top  \end{pmatrix}^\top$, where $\bm \delta_k$ is the vector of all standardized effects $\theta(j,k)$ of trial $k$ with IPD, i.e. $j=1,\ldots, q$ and $k=z+1,\ldots, q$; and $\bm \beta_k=\begin{pmatrix} \bm \beta_{1k}^\top & \ldots & \bm \beta_{jk}^\top & \ldots & \bm \beta_{qk}^\top \end{pmatrix}^\top$ is the vector of all parameters indexing the $q$ nuisance propensity score models needed to estimate $\bm \delta_k$. Besides, denote $\bm \Psi_k (\bm \theta_k, \bm O)$ the estimating function of $\bm \theta_k$. When the moment-based approach (\ref{eq7.5}) with $\bm\phi(\bm L) = \bm L$ is used to estimate the propensity model (\ref{eqw}), the component of $\bm \Psi_k (\bm \theta_k, \bm O)$ corresponding to $\bm  \beta_{jk}$ and $\theta(j,k)$ can be expressed as:
\[\bm \Psi_{\bm\beta_{jk}}= I(S=k) \bm\phi(\bm L)\exp(\bm L^\top\bm \beta_{jk}) - \bm\phi(\bm L) I(S=j),\]
 \[\Psi_{\theta(j,k)} = I(S=k)~ Y [(r_{1k}^{-1} + r_{0k}^{-1})X-r_{0k}^{-1}] ~\exp(\bm \beta_{jk}^\top \bm L) - \theta(j,k),\] 
respectively. As $\E \{\bm\Psi_k(\bm\theta_k,\bm O)\} = \bm 0$, the asymptotic distribution of $\bm{\hat \theta}_k$ obtained by solving $\sum_{i=1}^n \bm \Psi_k(\bm{\hat\theta}_k,\bm O_i) = \bm 0$ can be dervied from M-estimation theory as:
\[\sqrt{n}(\bm{\hat\theta}_k - \bm \theta_k) \xrightarrow{d} \mathcal{N} \big [\bm 0, ~ \E \left( - \bm \partial \bm \Psi_k/\bm \partial \bm \theta_k\right)^{-1} \cdot\bm{\Sigma}_{\bm \Psi_k} \cdot\E \left( - \bm \partial \bm \Psi_k/\bm \partial \bm \theta_k\right)^{-1,\top} \big ]\]
where $\bm{\Sigma}_{\bm \Psi_k} =  \E \{\bm \Psi_k(\bm \theta_k, \bm O)~ \bm \Psi_k (\bm \theta_k, \bm O)^\top\}$ and $\xrightarrow{d}$ denotes convergence in distribution. To estimate the asymptotic variance of $\sqrt{n}(\bm{\hat\theta}_k - \bm \theta_k)$, one then needs to estimate $\E \{ \bm \partial \bm \Psi_k/ \bm \partial \bm \theta_k\}$ and $\bm{\Sigma}_{\bm \Psi_k}$. Notice that the matrix $\bm \partial \bm \Psi_k/ \bm \partial \bm \theta_k$ only includes components of the form $I(S=k) \cdot \bm g(X, Y, \bm L)$ for some function $\bm g(\cdot)$. Consequently, the sample analogue of $\E\{\bm \partial \bm \Psi_k/ \bm \partial \bm \theta_k\}$ can be computed by using the IPD of trial $k$ only. In contrast, in the Online Supplementary Materials, we show that to estimate $\bm{\Sigma}_{\bm \Psi_k}$, one will additionally need an estimate for the covariance matrix of $\bm L$ in trial $j=1,\ldots, z$, i.e. $\mathrm{cov}(\bm L\mid S=j)$, which is often unavailable in practice.

To overcome the above challenge, one simple approach is to assume that the correlation matrix of $\bm L$ is constant across different populations as in some previous proposals \citep{rott2024causally,phillippo2020multilevel}. This common correlation matrix can then be estimated by using the IPD of trials $k=z+1,\ldots, q$, which then allows one to obtain an estimate for $\cov(\bm L\mid S=j)$. This so-called \textit{correlation extrapolating approach}, however, is biased if the underlying assumption of equal correlation across populations is invalid. Alternatively, note that:
\begin{align}\label{we}
    \E(\bm L \bm L^\top \mid S=j)=\frac{1}{\P(S=j)}  \E\left\{I(S=k) \bm L \bm L^\top  \frac{\P(S=j\mid \bm L)}{\P(S=k\mid \bm L)}\right\}
\end{align}
An estimate $\hat \E_k(\bm L \bm L^\top \mid S=j)$ for $\E(\bm L \bm L^\top \mid S=j)$ can thus be obtained by solving the sample analogue of this equation, using the data from trial $k$ with available IPD (i.e. $z+1\le k \le q$) along with model (\ref{eqw}). A single estimate $\hat \E^*(\bm L \bm L^\top \mid S=j)$ can then be obtained by averaging $\hat \E_k(\bm L\bm L^\top \mid S=j)$. Such an approach is valid provided that the propensity score model (\ref{eqw}) is correctly specified. In a similar way, we can also obtain a weighting-based estimate $\hat \E^*(\bm L\mid S=j)$ for $\E(\bm L\mid S=j)$. The missing covariance matrix in population $j$ can then be calculated as: 
\[\hat{\cov}(\bm L\mid S=j)=\hat \E^*(\bm L \bm L^\top \mid S=j) - \hat\E^*(\bm L\mid S=j)\cdot \hat\E^{*,\top}(\bm L\mid S=j). \]
\begin{remark}
Although the first and second moments of $\bm L$ in trial $j$ are assumed to be available in the trial's report, re-estimating these by the proposed weighting-based approach ensures that the obtained estimate $\ecov(\bm L\mid S=j)$ of $\cov(\bm L\mid S=j)$ is positive-definite. Indeed, for any vector $\bm v\in {\rm I\!R}^{p-1}$, one has: \[\bm v^\top \bm{\hat\Sigma}_{L\mid j}\bm v = (k-z)^{-1}\sum_{k:k>z}n_j^{-1}\sum_{i:S_i=k} m(\bm L_i,\bm{\hat\beta}_{jk})\{(\bm L_i-\hat{\bm\mu}_{L\mid j})^\top \bm v\}^2 \ge 0\]
where $\bm{\hat\Sigma}_{L\mid j}=\ecov(\bm L\mid S=j)$ and $\hat{\bm\mu}_{L\mid j}=\hat\E^*(\bm L\mid S=j)$. When $\{\bm L_i - \hat{\bm\mu}_{L\mid j}\}$ span ${\rm I\!R}^{p-1}$ then $\bm v^\top \bm{\hat\Sigma}_{L\mid j}=0$ if and only if $v=0$. This implies that $\bm{\hat\Sigma}_{L\mid j}$ is positive-definite.
\end{remark}

Once $\cov(\bm L\mid S=j)$ is estimated, constructing a sandwich estimator for the asymptotic covariance matrix of $\bm{\hat\theta}_k$ is straightforward. The closed form expression of this estimator can be found in the Online Supplementary Materials. In practice, to avoid manual computation of this sandwich estimator, one user-friendly strategy is to simulate pseudo-IPD for each trial $j=1,\ldots,z$ such that the summary statistics of these pseudo-IPD match $\bm{\hat\Sigma}_j$ and other summary statistics available in trial $j$. The asymptotic covariance matrix of $\bm {\hat\theta}_k$ can then be obtained by applying built-in functions for sandwich estimators in standard statistical software on these pseudo-IPD and the real IPD of trial $k$ (table 1). Provided that the propensity score model (\ref{eqw}) is correctly specified, the obtained covariance matrix estimate of $\bm{\hat \theta}_k$ will be valid. Note that correct specification of the propensity score model is also required for the consistency of $\bm{\hat \theta}$, so the proposed procedure for quantifying uncertainties in fact does not require any additional assumptions. This is an advantage of our proposal compared to previously suggested methods.
\begin{table}[tbh!]
\centering
\caption{Estimating the asymptotic variance of the case-mix standardized treatment effects $\theta(j,k)$}
\scalebox{0.95}{\begin{tabular}{ p{0.5cm} p{14cm}}
\hline
Step & Procedure\\[2pt] \hline
1. & For $j=1,\ldots, z$, obtain an estimate $\ecov(\bm L\mid S=j)$ for $\cov(\bm L\mid S=j)$ by the proposed weighting-based approach (\ref{we}) along with model (\ref{eqw}).\\[2pt]
2. & Simulate pseudo-IPD for each trial $j=1,\ldots,z$ such that the summary statistics of these pseudo-IPD match $\ecov(\bm L\mid S=j)$ and other summary statistics assumed available in trial $j$.\\[2pt]
3 & Estimate the asymptotic covariance matrix of $\bm \theta$ by a sandwich estimator, using the pseudo-IPD from trial $j=1, \ldots, z$ and real IPD from trial $k=z+1,\ldots q$.\\[2pt]
\hline
\end{tabular}}
\end{table}

\subsection{A simulation study}
We now conduct a simulation study to assess the finite sample performance of the proposed approach of case-mix standardization. We generate hypothetical data for three randomized trials that evaluates the effect of a binary treatment $X$ on a binary outcome $Y$ on the risk difference scale. The trial populations are heterogeneous in the distribution of a continuous variable $L_1$ and a binary variable $L_2$. We consider two settings. In the first setting, the individual-level data vector $O=(S, L_1, L_2, X, Y)$ of $n$ patients is generated by the following mechanism:
\begin{align*}
    L_1 &\sim Unif(0,1)\\
    L_2 &\sim Bin(0.5)\\
    X &\sim Bin(0.5)\\
    P(S=s\mid L) &=\frac{\mathbb{1}_{s=1}+ \mathbb{1}_{s=2}~\mathrm{exp}(1 - L_1 - L_2) + \mathbb{1}_{s=3}~\mathrm{exp}(-1 + L_1 + L_2)}{1+\mathrm{exp}(1 - L_1 - L_2) + \mathrm{exp}(-1 + L_1 + L_2)} \\
    P(Y=1\mid X, L, S) &= \mathrm{expit}\{-0.25 + (1.75\mathbb{1}_{s=1} + 0.5\mathbb{1}_{s=2} - 0.25\mathbb{1}_{s=3})X - 1L_2 + 1L_1 - 2XL_2 + 2XL_1\}
\end{align*}
where $\mathbb{1}_{s=s^*}=1$ if $s=s^*$ and $0$ otherwise, with $s,s^*=1,2,3$. The unstandardized average treatment effects across populations are heterogeneous, i.e., $\theta(1,1) = 0.27$; $\theta(2,2) = 0.15$ and $\theta(3,3) = -0.07$, due to the difference between studies in the distribution of $(L_1,L_2)$ and in the coefficient of the treatment in the underlying outcome model (i.e. beyond case-mix heterogeneity).
In the second setting, we generate $L_1,L_2,X,Y$ by using the same models as in setting 1, but we generate $S$ by an alternative model:
\begin{align*}
    P(S=s\mid L) &=\frac{\mathbb{1}_{s=1}+ \mathbb{1}_{s=2}~\mathrm{exp}(1 - L_1 - L_2 - 3L_1^2) + \mathbb{1}_{s=3}~\mathrm{exp}(-1 + L_1 + L_2 - 3L_1^2)}{1+\mathrm{exp}(1 - L_1 - L_2 - 3L_1^2) + \mathrm{exp}(-1 + L_1 + L_2 - 3L_1^2)} 
\end{align*}
The unstandardized treatment effects in setting 2 are $\theta(1,1) = 0.24$; $\theta(2,2) = 0.15$ and $\theta(3,3) = -0.15$.

In both settings, we assume that the individual-level data of trial $S=1$ is not accessible. However, the sample mean and sample variance of $L_1$ and $L_2$ in this trial are available. The inverse weighting approach proposed in section 2 is then used to transport results of trial $S=2$ and $S=3$ over the case-mix of the target population of trial $S=1$, assuming a multinomial logistic weight model with main effects of covariates $L_1$ and $L_2$. This weight model is thus correctly specified in setting 1 but mispecified in setting 2, due to the omission of the quadratic term $L_1^2$. To estimate the covariance matrix of $(L_1,L_2)$ in trial $1$, we consider the weighting approach based on (\ref{we}). For each standardized effect $\theta(1, k)$ where $k=2,3$, we evaluate the average of the difference $\hat \theta(1, k) - \theta(1,k)$, the median of the variance estimate $\hat{\V}\{\hat\theta(1,k)\}$, the mean squared error (MSE) of $\hat\theta(j,k)$ and the coverage of the 95\% confidence interval constructed for $\theta(j,k)$ across simulations. The true value of $\theta(1,k)$ is computed based on a sample of $10^7$ patient profiles, and the true value of var$\{\hat\theta(1,k)\}$ is computed based on the sample variance of $\hat\theta(j,k)$ obtained across all simulations. We implement $5000$ simulations in each setting, and consider three sample sizes, i.e. $n = 500, 1000$ and $2000$.

\begin{table}[]
    \centering
    \captionsetup{width=.75\textwidth}
    \caption{Simulation study 1. Bias: the averaged difference between $\hat\theta(j,k)$ and $\theta(j,k)$ across simulation. var$\{\hat\theta(j,k)\}$: the variance of $\hat\theta(j,k)$ estimated across simulation. $\hat{\mathrm{var}}\{\hat\theta(j,k)\}$: the median of the variance estimate of $\hat\theta(j,k)$ across simulation. MSE: Mean squared error of $\hat\theta(j,k)$. Coverage: coverage of the 95\% confidence interval constructed for $\hat\theta(j,k)$.}
    \scalebox{0.8}{\begin{tabular}{cccccccc}
    \hline
    Setting & Parameter & Sample size & Bias & var$\{\hat\theta(1,k)\}$ & $\hat{\mathrm{var}}\{\hat\theta(1,k)\}$ & MSE& Coverage (\%)\\
    \hline
    1& $\theta(1,2)$ & 500 & $-1.7\times 10^{-3}$& $1.37\times 10^{-2}$ & $1.29\times 10^{-2}$ & $1.36\times 10^{-2}$ & 94.5\\
    & $\theta(1,3)$ & 500 & $-5.1\times 10^{-3}$& $1.64\times 10^{-2}$& $1.51\times 10^{-2}$ & $1.65\times 10^{-2}$& 93.2\\
    & $\theta(1,2)$ & 1000 & $-3.6\times 10^{-3}$ & $6.49\times 10^{-3}$& $6.60 \times 10^{-3}$ & $6.51\times 10^{-3}$ & 95.3\\
    & $\theta(1,3)$ & 1000 & $1.9\times 10^{-3}$& $7.06\times 10^{-3}$& $7.63\times 10^{-3}$& $7.06\times 10^{-3}$& 95.9\\
    & $\theta(1,2)$ & 2000 & $0.2\times 10^{-3}$ & $1.25\times 10^{-3}$ & $1.31\times 10^{-3}$ & $1.25\times 10^{-3}$& 96.1\\
    & $\theta(1,3)$ & 2000 & $2.4\times 10^{-3}$ & $1.51\times 10^{-3}$ & $1.53\times 10^{-3}$ & $1.51\times 10^{-3}$ & 94.4\\
    \hline
    2&  $\theta(1,2)$ & 500 & $-0.9\times 10^{-2}$ & $5.73\times 10^{-2}$ & $4.10\times 10^{-2}$ & $5.73\times 10^{-2}$ & 89.1\\
    &  $\theta(1,2)$ & 500 & $-0.2\times 10^{-2}$ & $4.51\times 10^{-2}$ & $3.45\times 10^{-2}$ & $4.51\times 10^{-2}$ & 90.6\\
    &  $\theta(1,2)$ & 1000 & $-1.0\times 10^{-2}$ & $2.84\times 10^{-2}$ & $2.35\times 10^{-2}$ & $2.85\times 10^{-2}$ & 91.1\\
    &  $\theta(1,2)$ & 1000 & $0.7\times 10^{-2}$ & $1.96\times 10^{-2}$ & $1.87 \times 10^{-2}$ & $1.96\times 10^{-2}$ & 93.6\\
    &  $\theta(1,2)$ & 2000 & $0.5\times 10^{-2}$ & $1.45\times 10^{-2}$ & $1.27\times 10^{-2}$ & $1.45\times 10^{-2}$ & 92.6\\
    &  $\theta(1,2)$ & 2000 & $0.3\times 10^{-2}$ & $1.02\times 10^{-2}$ & $0.97\times 10^{-2}$ & $1.02\times 10^{-2}$ & 92.7\\
    \hline
    \end{tabular}}
    \label{tab:sim1}
\end{table}

Results of this simulation study can be found in table \ref{tab:sim1}. In setting 1, the proposed approach provides consistent estimates for both parameters $\theta(1,2)$ and $\theta(1,3)$. The suggested procedure for quantifying uncertainties also return valid estimates for the asymptotic variances of the standardized effect estimates. The 95\% confidence intervals of both parameters slightly under-cover when the sample size is small. However, this issue quickly vanishes when the sample size becomes sufficiently large. 

In setting 2, mispecifying the weight model leads to biased estimates for $\theta(1,2)$ and $\theta(1,3)$, incorrect variance estimates and under-covering 95\% confidence intervals. These problems are improved at larger sample sizes, but the improvement is slower than when the weight model is correctly specified.
\section{Causally interpretable meta-analysis models}

Standard pairwise meta-analysis often focuses on averaging results of different trials, while overlooking the target patient population for which the obtained average describes the treatment effect.
Causal meta-analysis overcomes this drawback by considering case-mix standardization prior to evidence synthesis. For instance,  results of different trials $k=1,\ldots, q$ are transported to a target population $j$ to obtain the estimates $\hat \theta(j,k)$. These estimates are then meta-analyzed to derive a summary statistic that expresses the treatment effect for population $j$, while the between-trial variance from such a meta-analysis expresses how much results from different trials vary even when being hypothetically conducted in the same population (i.e. beyond case-mix heterogeneity).
One shortcoming of this approach, however, is that one separate meta-analysis needs to be conducted for each target population $j$. 
Alternatively, we propose a random-effect model that simultaneously meta-analyzes all case-mix standardized effect estimates $\hat\theta(j,k)$. The summary treatment effect for each trial population $j=1,\ldots,q$ can then be calculated once the model is fitted. Such a model assumes:
\begin{equation}\label{eq6}
\begin{aligned}
\hat\theta(j,k) &= \theta(j,k)+ \varepsilon_{jk}; \\
\theta(j,k) &=\theta + \beta_j + \gamma_k; \\
\beta_j &\sim \mathcal{N}(0,\omega^2);  \\
\gamma_k &\sim \mathcal{N}(0, \xi^2);\\ 
\varepsilon_{jk} &\sim \mathcal{N}(0, \sigma_{jk}^2);\\ 
\beta_j &\independent\gamma_k \independent \varepsilon_k. 
\end{aligned}
\end{equation}
for $j,k=1,\ldots,q$. Here, $\beta_j$ and $\gamma_k$ denote two independent mean-zero random effects with variance $\omega^2$ and $\xi^2$, respectively; and $\epsilon_{jk}$ denotes a mean-zero random error with variance $\sigma^2_{jk}$. As in standard random-effect meta-analysis models, we here assume that the available studies are a representative sample from infinitely many studies that could have been conducted. Two random effects are then included to capture the variability of $\theta(j,k)$ when $j$ varies and $k$ varies (see below). The proposed model (\ref{eq6}) can thus be viewed as an extension of standard meta-analysis model to summarize case-mix standardized effect estimates, in the sense that when this model holds, the standard random-effect meta-analysis model proposed in the literature to summarize $\hat\theta(j,j)$ also holds.

Quite intuitively, $\sigma^2_{jk}= \V\{\hat\theta(j,k)\mid \theta(j,k)\}$ reflects the uncertainty due to random chance when estimating $\theta(j,k)$ by $\hat\theta(j,k)$. Beyond random chance, we have non-random case-mix and beyond case-mix heterogeneity (see equation (\ref{eq2}) and remark 1). In principle, when different trials are conducted in the same target population, case-mix heterogeneity is avoided but beyond case-mix heterogeneity may still present due to the difference in trial protocols. In contrast, when the same trial is conducted (by exactly the same protocol) in different target populations, beyond case-mix heterogeneity is avoided but case-mix heterogeneity may still present. Under model (\ref{eq6}), the variability observed among trial results when all trials are conducted in the same population can be expressed as: 
\[\V\{\theta(j,k)\mid j\} = \xi^2\] 
where $\V\{\theta(j,k)\mid j\}$ denotes the variance of $\theta(j,k)$ when $j$ is fixed and $k$ varies. This suggests that $\xi^2 = \V(\gamma_k)$ reflects so-called beyond case-mix heterogeneity. In contrast, the variability observed in the results of a trial when it is conducted (with the same protocol) in different populations can be expressed as:
\[\V\{\theta(j,k)\mid k\}=\omega^2\] 
where $\V\{\theta(j,k)\mid k\}$ denotes the variance of $\theta(j,k)$ when $k$ is fixed and $j$ varies. This suggests that $\omega^2 = \V(\beta_j)$ reflects so-called case-mix heterogeneity. Finally, note that: 
\[\V\{\theta(j,j)\}=\omega^2 + \xi^2\] 
which suggests that the total heterogeneity $\tau^2 := \V\{\theta(j,j)\}$ often reported in standard random-effect meta-analysis can be perfectly decomposed into case-mix and beyond case-mix heterogeneity, i.e. $\tau^2 = \omega^2 + \xi^2$. The population $j_0$--specific summary treatment effect can also be computed from model (\ref{eq6}) as $\hat\theta + \hat \beta_{j_0}$, where $\hat\theta$ is a consistent estimate of the fixed-effect component $\theta$ and $\hat \beta_{j_0}$ is an estimate of the random effect $\beta_{j_0}$, where $j_0 = 1,\ldots,q$. 

\begin{table}[!htb]
\captionsetup{width=.75\textwidth}
    \caption{Missingness of $ {\hat \theta}(j,k)$ due to individual-level data restriction. $q$: total number of studies. $z$: total number of studies with aggregated data. The value in each cell is $1$ if $\theta(j,k)$ is estimable by the inverse weighting approach proposed in section 2, and $0$ otherwise.}
    \centering
\begin{TAB}(e,1cm,1cm){|c|c|c|c|c|c|c|c|}{|c|c|c|c|c|c|c|c|}
\diagbox[]{$\bm j$}{$\bm k$}  & $\bm 1$ & $\bm 2$ & $\bm\ldots$ & $\bm z$ & $\bm{z+1}$ & $\bm\ldots$ & $\bm q$ \\
$\bm 1$ & 1 & 0 & 0 & 0  & 1 & 1 & 1 \\
$\bm 2$ & 0 & 1 & 0 & 0  & 1 & 1 & 1 \\
$\bm \ldots$ & 0 & 0 & 1 & 0 & 1 & 1 & 1 \\
$\bm z$  &  0 & 0 & 0 & 1 & 1 & 1 & 1 \\
$\bm{z+1}$ & 0 & 0 & 0 & 0 & 1 & 1 & 1 \\
$\bm\ldots$ & 0 & 0 & 0 & 0 & 1 & 1 & 1 \\
$\bm q$ &0 & 0 & 0 & 0 & 1 & 1 & 1 \\
\end{TAB}
\label{tab2}
\end{table}

When all trials in the meta-analysis have accessible individual-level data, one can obtain $q^2$ estimates $\hat\theta(j,k)$ for $j,k=1,\ldots, q$. Fitting model (\ref{eq6}) on these $q^2$ estimates is feasible by restricted maximum likelihood or Markov Chain Monte Carlo (MCMC) simulations. In contrast, in the current context of multiple aggregated data trials, not all $q^2$ parameters $ { \theta}(j,k)$ can be estimated and subsequently meta-analyzed. Indeed, the inverse weighting approach proposed in section 2 only allows one to standardize results of a trial with individual-level data over the target population of another trial with aggregated data, and not vice versa. 
Fortunately, the parameters indexing model (\ref{eq6}) can still be indirectly estimated by using available $\hat\theta(j,k)$. This is possible upon noting that for a given trial $k$ with IPD, its results can be transported to any other population $j=1,\ldots,q$, regardless of whether the IPD is available  in such population (table \ref{tab2}). Hence, the case-mix heterogeneity $\omega^2$ can be validly estimated by summarizing the standardized effects $\hat\theta(j,k)$ of the same trial $k$. Similarly, the total heterogeneity $\tau^2$ can be estimated by summarizing the original results $\hat\theta(j,j)$ that are available for all trials. 

The above arguments suggest that the following submodel can be fitted on the available $\hat\theta(j,k)$:
\begin{equation}\label{eq7}
    \begin{aligned}
    \hat\theta(j,k) &= \theta_k + \beta_j + \epsilon_{jk} & 
    \quad \hat\theta(j,j) &= \theta + (\beta_j + \gamma_j) + \epsilon_{jj} \\
    \beta_j &\sim \mathcal{N}(0,\omega^2) & \quad 
    \beta_j + \gamma_j &\sim \mathcal{N}(0, \tau^2)  \\
    \epsilon_{jk} &\sim \mathcal{N}(0,\sigma^2_{jk}) &\quad \epsilon_{jj} &\sim \mathcal{N}(0,\sigma^2_{jj})
    \end{aligned}
\end{equation}
for $ j=1,\ldots, q; k = z+1,\ldots, k$. Here, $\theta_k$ denotes the summary treatment effect obtained from meta-analyzing $\hat\theta(j,k)$ with $k$ fixed among $z+1,\ldots, q$, and $j$ varying from $1$ to $q$. Fitting this submodel can then provide valid estimates $\hat\theta$, $\hat\omega^2$ and $\hat\tau^2$ for $\theta,\omega^2$ and $\tau^2$ indexing model (\ref{eq6}), respectively. The beyond case-mix variance $\xi^2$ can be indirectly estimated as $\hat\xi^2 = \max(0,\hat\tau^2-\hat\omega^2)$. Finally, a summary estimate for each trial population $j$ can be obtained by $\hat\theta + \hat\beta_j$ for $j=1,\ldots, q$, where $\hat\beta_j$ is an estimate of $\beta_j$ also obtained from the proposed submodel. For instance, when the submodel is fitted by restricted maximum likelihood, $\hat\beta_j$ is the best linear unbiased prediction (BLUP) of $\beta_j$. In contrast, when the submodel is fitted by MCMC simulations, one can take random samples from the posterior distribution of $\theta + \beta_j$ and summarize these samples by median and 95\% credible interval.

\begin{remark}
    The estimation procedure (\ref{eq7}) relies on the constant variance assumption encoded in model (\ref{eq6}), which states that the case-mix and beyond case-mix heterogeneity variances are constants across $j$ and $k$, respectively. In other words, model (\ref{eq6}) assumes that the variance of elements in $\mathcal{S}_j=\{\theta(j,k) \mid k \in \mathbb{N}\}$  is unchanged across different sets $\mathcal{S}_j$ and equals $\xi^2$, for $j\in \mathbb{N}$; while the variance of elements in $\mathcal{S}_k=\{\theta(j,k) \mid j\in \mathbb{N}\}$ is unchanged across different sets $\mathcal{S}_k$ and equals $\omega^2$, for $k\in \mathbb{N}$. With the current dataset at hand, we have a random subset of elements for each $\mathcal{S}_k$ with $k=z+1,\ldots,q$, so the common variance $\omega^2$ is estimable by the sample variance of elements within each of these subsets.
\end{remark}

In what follows, we conduct a simulation study to illustrate the use of the proposed strategy to estimate the random-effect model (\ref{eq6}). Assume that we want to meta-analyze the standardized effect estimates obtained from $q\in \{20,40\}$ trials, among which $z=10$ studies have accessible individual-level data. To generate $\hat{\theta}(j,k)$ where $j,k=1,\ldots, q$, we consider the random-effect model (\ref{eq6}) with $\theta = 0$ and $\xi^2 = \omega^2$ = 0.5. The total heterogeneity $\tau^2$ thus equals 1. To generate the residual covariance matrix $\bm\Sigma$, we consider the simulation method suggested by \citet{vo2021assessing}. More details on $\bm\Sigma$ are also provided in the Online Supplementary Materials. 

\begin{figure}[H]
    \centering
    \includegraphics[scale = 0.65]{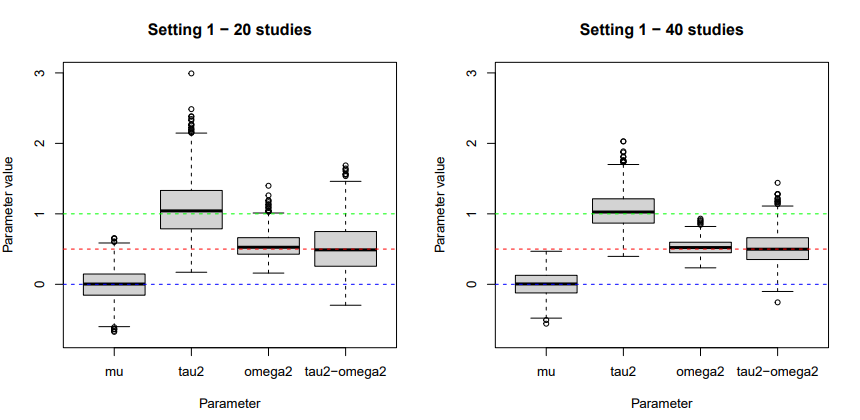}
    \captionsetup{margin=2cm}
    \caption{Estimation of the two-random-effect model (\ref{eq6}) with $\theta=0,\tau^2 = 1, \omega^2 =  0.5, \xi^2 = 0.5$. In each graph, we provide the distribution of the posterior median of $\theta,\omega^2$, $\tau^2$ and $\xi^2 = \tau^2 - \omega^2$. The dotted lines in blue, red and green represent the true value of these parameters in each setting, respectively.}
    \label{fig:1}
\end{figure}
Parameters $\theta, \xi^2$ and $\tau^2$ in the random-effect model (\ref{eq6}) are estimated by fitting the submodel (\ref{eq7}) by Markov Chain Monte Carlo (MCMC) simulations. As in many other meta-analysis proposals, we additionally assume that the residual covariance matrix $\bm\Sigma$ is a priori known across all settings to simplify the estimation procedure. In each simulation, we specify weakly informative, independent prior distributions for the unknown parameter $\theta, \xi^2$ and $\tau^2$. More precisely, the prior for $\theta$ is the normal distribution $\mathcal{N}(0,1\,000)$, and for $\xi^2$ and $\tau^2$ it is the uniform distribution $U(0,100)$. The MCMC simulations are implemented by the function \texttt{jags.model()} from the \texttt{R} package \texttt{rjags}, with 2 parallel chains and 10000 iterations for adaptation. A sample of 1000 observations per chain (with a thinning interval of 5) is then taken from the posterior distribution of each parameter, i.e. $\theta, \xi^2,\tau^2$ and $\tau^2 - \xi^2$, using function \texttt{coda.samples()} from \texttt{rjags}. The median of the posterior sample is recorded as the final estimate of each parameter. We implement 1000 simulations in each setting.

Results of this simulation study can be found in figure \ref{fig:1}. In both settings, the posterior median of each parameter (among $\theta,\omega^2,\tau^2$ and $\xi^2 = \tau^2 - \omega^2$) is close to its true value. This suggests that the proposed procedure to estimate model (\ref{eq6}) is theoritically valid, provided that this model correctly describes the distribution underlying the standardized treatment effects $\theta(j,k)$.

\section{Application on real data}
We apply the proposed method to meta-analyze two multi-center randomized studies, UltIMMa-1, and UltIMMa-2 \citep{gordon2018efficacy}, which compare the clinical effectiveness of risankizumab and ustekinumab among patients with moderate to severe psoriasis. The outcome of interest in both studies is a binary variable that indicates whether a patient achieves an improvement of at least 75\% in the psoriasis area and severity index (PASI) score at week 12 compared to baseline. For illustrative purposes, we consider different centers in these two trials as independent studies. For instance, we consider the Austrian and Japanese centers of UltIMMa-1 as study 1, the Canadian and Korean centers of UltIMMa-1 as study 2, the American center of UltIMMa-1 as study 3, the Canadian center of UltIMMa-2 as study 4 and the American center of UltIMMa-2 as study 5. Data on four covariates are collected across five studies, which include gender, age, PASI score at baseline and body mass index (BMI) at baseline (table \ref{tab:4}).

We assume that that the IPD are available in trials 3, 4, and 5; and the aggregated data (i.e. points (i)-(v) of section 2.2) are available in trials 1 and 2. A multinomial logistic model that includes all baseline covariates as main effects is assumed to describe the relationship between the trial membership $S$ and the baseline covariates $L$. To standardize results of a trial $k$ with IPD $(k=3,\ldots, 5)$ over the case-mix of another trial $j$ with or without IPD $(j=1,\ldots, 5)$, the moment-based approach relying on equation (\ref{eq6}) with $\bm\phi(\bm L) = \bm L$ is used to estimate the weight $\hat\P(S_i=j\mid \bm L_i)/\hat\P(S_i=k\mid \bm L_i)$ for each individual $i$ in trial $k$. We truncate the weights at the $95^{th}$ percentile prior to case-mix standardization. 

To establish the covariance matrix of the standardized treatment effects $\hat\theta(j,k)$, we first estimate cov$(\bm L\mid S=j)$ for $j=1,2$ by applying the weighting approach (\ref{we}) on the IPD of trials $k=3,\ldots,5$. Pseudo-IPD for trial $j=1,2$ are then generated such that the summary statistics of these pseudo-IPD match $\hat{\cov}(\bm L\mid S=j)$ and other summary statistics available in trial $j$. A sandwich estimator for the variance of $\{\hat\theta(j,k):j=1,2;~k=3,\ldots,5\}$ can then be obtained from the pseudo-IPD of trial $j=1,2$, and the real IPD of trials $k=3,\ldots, 5$. Finally, model (\ref{eq6}) is used to summarize all case-mix-standardized estimates. This model is estimated by fitting the submodel (\ref{eq7}) by Markov Chain Monte Carlo simulations. For this step, we use the same setups as in Section 4.2 for the prior distribution of each parameter. The summary treatment effect in each trial population is then derived by summarizing the posterior distribution of $\theta + \beta_j$, where $j=1,\ldots, 5$, by using the median and the 2.5 and 97.5\% quantiles. The \texttt{R} code for implementation of this analysis can be found on \href{https://github.com/votatthang/causalmeta}{https://github.com/votatthang/causalmeta}.

\begin{figure}[tbh!]
    \centering
    \includegraphics[width=0.6\linewidth]{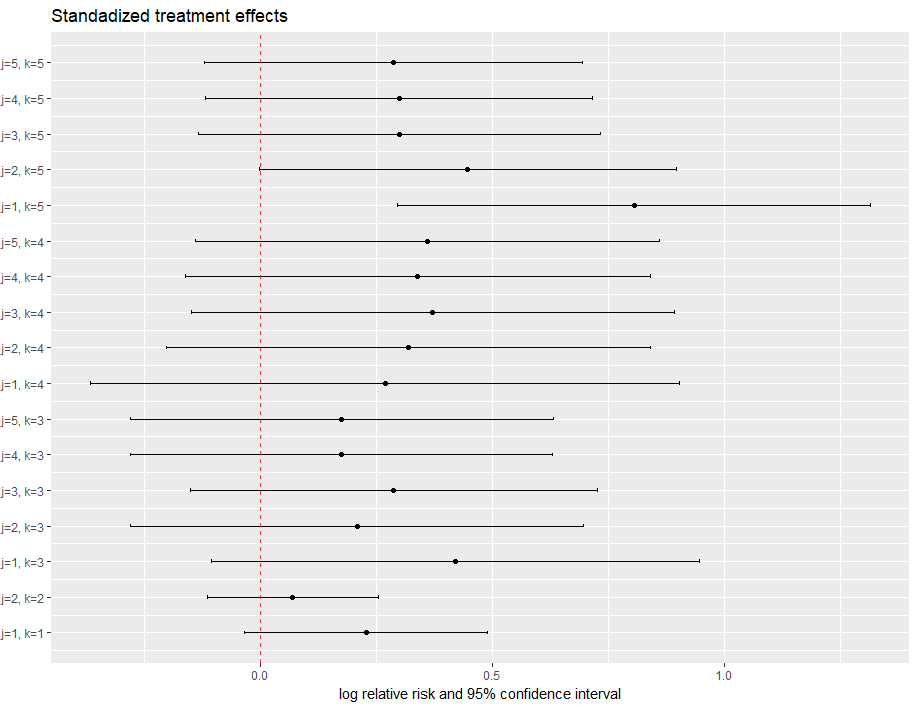}
    \captionsetup{margin=2cm}
    \caption{Data analysis: standardized treatment effects across studies. Each horizontal line on the forest plot represents the estimate and corresponding 95\% confidence interval of the standardized treatment effect $\theta(j,k) = \log \{P(x=1,k)=1\mid S=j\} - \log \{P(x=0,k)=1\mid S=j\}$, where $x=1$ denotes risankizumab and $x=0$ denotes ustekinumab. Individual participant data are available in studies $k=3,4,5$.}
    \label{fig:2}
\end{figure}
\begin{figure}[tbh!]
    \centering
    \captionsetup{margin=2cm}
    \includegraphics[width=0.6\linewidth]{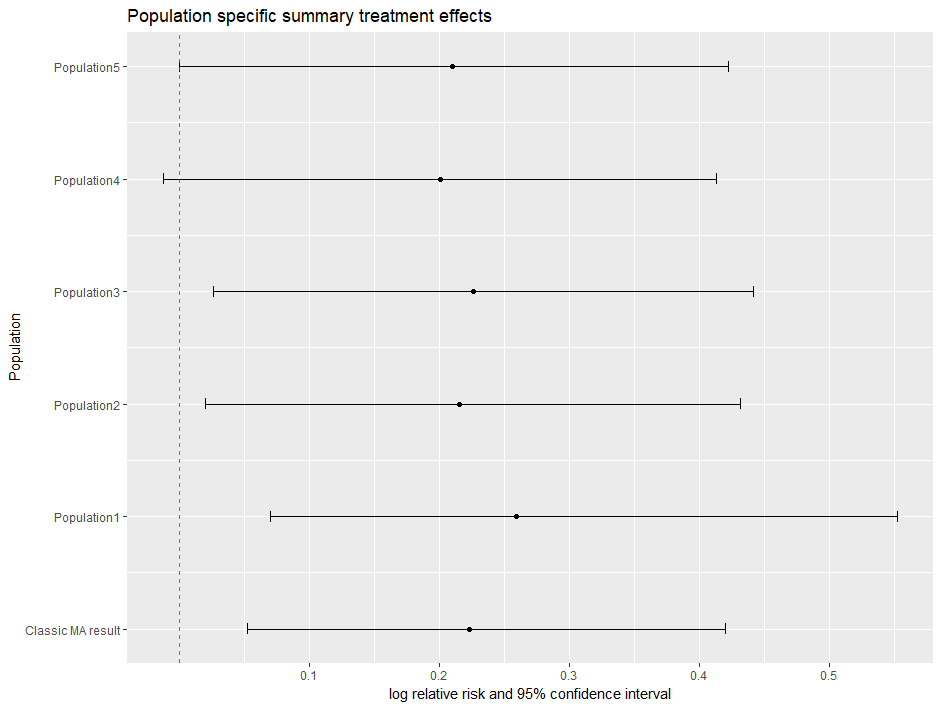}
    \caption{Data analysis: population-specific summary treatment effects.  Each horizontal line on the forest plot represents the median and corresponding 95\% credible interval of 1000 random samples taken from the posterior distribution of the population $j-$specific summary treatment effect defined on the log relative risk scale (risankizumab versus ustekinumab). This distribution is obtained by using Markov Chain Monte Carlo simulations to indirectly estimate model (\ref{eq6}) via the submodel (\ref{eq7}).}
    \label{fig:3}
\end{figure}
Results of this illustrative meta-analysis are provided in figure 2 and 3. By fitting model (5), we obtain a summary treatment effect $\hat\theta$ of $0.223$ (95\% credible interval (CI) $ 0.052; 0.420$), a total heterogeneity variance $\hat\tau^2$ of $9.8\times 10^{-3}$ (95\%CI: $1.91\times 10^{-5}$; $0.204$), and a case-mix heterogeneity variance $\hat\omega^2$ of $1.2\times 10^{-3}$ (95\%CI: $1.38\times 10^{-6}; 0.060$). This expresses that the difference between studies in the distribution of the baseline covariates $L$ accounts for about 12\% of the total heterogeneity observed across trial results. The population-specific summary effects are also visualized in figure 3, which shows better effectiveness of risankizumab compared to ustekinumab in all trial populations. Statistical significance is also obtained in population $S=1$ ($0.259$, 95\%CI: $0.070, 0.552$), population $S=2$ ($0.215$, 95\%CI: $0.020, 0.432$), population $S=3$ ($0.226$, 95\%CI: $0.026, 0.442$), and population $S=5$ ($0.210$, 95\%CI: $1.66\times 10^{-5}, 0.422$); but not in population $S=4$ ($0.201$, 95\%CI: $-0.012, 0.413$).

\section{Conclusion}
In this paper, we extend the moment-based methods frequently used for population adjustment in HTA to allow for the integration of aggregated data into a causally interpretable meta-analysis. The standardized effect estimates obtained by inverse weighting are summarized by a random-effect model which decomposes the total heterogeneity across trials into so-called case-mix and beyond case-mix heterogeneity. Such a model allows one to derive multiple summary treatment effect estimates that are specific for different trial populations in the meta-analysis. 

The strength and weakness of our proposed approach compared to other strategies suggested in the literature are briefly summarized in table \ref{tab:4}. To transport the treatment effect from a source population with IPD to a target population with aggregated data, previous methods often start by generating the pseudo-IPD for the covariates in the target population. This step requires a structural assumption on the covariance/correlation matrix of the covariates across populations, since information on such matrix is often not available in the aggregated data of the target population. Standard methods for case-mix standardization such as G-computation and inverse weighting are then applied on the IPD of the source trial and on the pseudo-IPD of the target population. When the outcome model and/or the propensity score model in these methods are correctly specified, bias may still arise if the assumption on the covariance/correlation matrix of covariates is violated, i.e. if the pseudo-IPD are not generated from the true data generating mechanism of the target population. Unlike these methods, our proposal does not require any assumption on the covariance of covariates in the target population. Alternatively, we estimate it by using the available IPD, along with the same weight model that is used for the next step of case-mix standardization. This helps reduce the number of assumptions required for valid standardized estimates.

\begin{table}[tbh!]
    \centering
    \captionsetup{width=.8\textwidth}
    \caption{\small Comparing different methods to combine IPD and aggregated data in causal meta-analysis. IPD: individual participant data. OCM: outcome model. PSM: propensity score model. FEM: Fixed-effect model. REM: Random-effect model.}
    \scalebox{0.7}{\rowcolors{2}{white}{gray!25}
    \begin{tabular}{p{8cm}p{5cm}p{4.7cm}p{4.5cm}}
    \hline
    \textbf{Method} & \citet{rott2024causally} & 
    \makecell[l]{\citet{remiro2022parametric}} &
    Our proposal\\
    \hline
    \multicolumn{4}{l}{\textbf{Case-mix standardization step}}\\
    \makecell[l]{- Additional assumption on the covariance/\\~~~correlation matrix of covariates} & Required & Required & Not required\\
    - Modeling assumption & \makecell[l]{Depends on the method\\ used for standardization, i.e.\\
    OCM if G-computation; \\PSM if inverse weighting}
    &\makecell[l]{ OCM (standardization\\by G-computation)} & \makecell[l]{ OCM (standardization\\by inverse weighting)}\\
    - Pseudo-IPD generation & Required & Required & \makecell[l]{Not required, but \\can be used to \\facilitate the
    estimation \\of the asymptotic variance}\\
    \makecell[l]{- Taking into account the \\~~ uncertainty due to generated pseudo-IPD} & By multiple imputations & By multiple imputations &  Not required\\
    - Potential risk under lack of case-mix overlap & \makecell[l]{Erroneous extrapolation \\(G-computation) \\or extreme weights \\(inverse weighting)} & Erroneous extrapolation & Extreme weights\\
    \hline
    \multicolumn{4}{l}{\textbf{Meta-analysis step}}\\
    - Model to combine standardized effects & Classical FEM/REM & Classical FEM/REM & \makecell[l]{Novel REM with\\ two random effects}\\
    - Target population & External or internal &External or internal & Internal\\
    - Extension to network of treatments &Possible & Available & Possible \\
    - Analysis repeated for different target populations & Required & Required & Not required\\
    \hline
    \end{tabular}
    \label{tab:5}}
\end{table}

Our proposal also has certain limitations that can be addressed in future works. First, case-mix standardization based on inverse weighting is known to have limited power. To overcome this, the proposed method can be ameliorated by adding into the estimating function of inverse weighting a mean-zero residual that further exploits the outcome information available in the source trial. Second, extreme weights may arise when there is limited overlap between the source trial and the target population.
While alternative approaches based on G-computation have been proposed in the literature to overcome this concern, these methods come with the risk of erroneous extrapolation that is difficult to control, and moreover can be very sensitive to the (outcome) model restrictions on effect modification. Besides, G-computation methods require the generation of pseudo-IPD in the estimation process, which---as we argue above---may easily lead to bias if the model used for pseudo-IPD generation is incorrect \citep{remiro2022parametric,vo2023cautionary}. Third, our proposal focuses mostly on deriving the treatment effect for the target population of each trial in the meta-analysis. In practice, the actual interest may lie in the treatment effect of an external target population that is different from that of eligible trials. Generalizing the proposed method to such a setting is important. Fourth,
a different set of baseline covariates may be collected in different trials \citep{vuong2024development}. Addressing this concern along with limited IPD access could further facilitate the use of causal meta-analysis in practice. Finally, the inclusion of multiple random effects to reflect different sources of heterogeneity can be generalized to models of network meta-analysis, to allow for the indirect comparison of multiple treatment options in different trial populations in the network. 

\section*{Acknowledgment}
Tat-Thang Vo is supported by the French National Research Agency (Agence Nationale de la Recherche), through a funding for Chaires de Professeur Junior (23R09551S-MEDIATION). All authors declare that they have no conflicts of interest.

\section*{Data availability statement}
The clinical trial data that support the findings in this paper are from the Vivli platform. More information on how to get access to these data is available on the platform’s website (\href{https://vivli.org/}{https://vivli.org/}). 
\bibliographystyle{plainnat}
\bibliography{refs}
\newpage
\appendix
\singlespacing
\section*{Online Supplementary Materials}
\addcontentsline{toc}{section}{Appendices}
\renewcommand{\thesubsection}{\Alph{subsection}}
\subsection{Statistical inference when transporting a trial with IPD to multiple target populations with IPD or aggregated data}

Denote $\bm \theta_k=\begin{pmatrix} \bm \beta_k^\top & \bm \delta_k^\top  \end{pmatrix}^\top$, where $\bm \delta_k$ is the vector of all standardized effects $\theta(j,k)$ of trial $k$ with IPD, i.e. $j=1,\ldots, q$ and $k=z+1,\ldots, q$; and $\bm \beta_k=\begin{pmatrix} \bm \beta_{1k}^\top & \ldots & \bm \beta_{jk}^\top & \ldots & \bm \beta_{qk}^\top \end{pmatrix}^\top$ is the vector of all parameters indexing the $q$ nuisance propensity score models needed to estimate $\bm \delta_k$. Besides, denote $\bm \Psi_k (\bm \theta_k, \bm O)$ the estimating function of $\bm \theta_k$. When the proposed moment-based approach with $\bm\phi(\bm L) = \bm L$ is used to estimate the propensity model (\ref{eqw}), the component of $\bm \Psi_k (\bm \theta_k, \bm O)$ corresponding to $\bm  \beta_{jk}$ and $\theta(j,k)$ can be expressed as:
\[\bm \Psi_{\bm\beta_{jk}}= I(S=k) \bm\phi(\bm L)\exp(\bm L^\top\bm \beta_{jk}) - \bm\phi(\bm L) I(S=j),\]
 \[\Psi_{\theta(j,k)} = I(S=k)~ Y [(r_{1k}^{-1} + r_{0k}^{-1})X-r_{0k}^{-1}] ~\exp(\bm \beta_{jk}^\top \bm L) - \theta(j,k),\] 
respectively. As $\E \{\bm\Psi_k(\bm\theta_k,\bm O)\} = \bm 0$, the asymptotic distribution of $\bm{\hat \theta}_k$ obtained by solving $\sum_{i=1}^N \bm \Psi_k(\bm{\hat\theta}_k,\bm O_i) = \bm 0$ can be dervied from M-estimation theory as:
\[\sqrt{n}(\bm{\hat\theta}_k - \bm \theta_k) \xrightarrow{D} \mathcal{N} \big [\bm 0, ~ \E \left( - \bm \partial \bm \Psi_k/\bm \partial \bm \theta_k\right)^{-1} \cdot\bm{\Sigma}_{\bm \Psi_k} \cdot\E \left( - \bm \partial \bm \Psi_k/\bm \partial \bm \theta_k\right)^{-1,\top} \big ]\]
where $\bm{\Sigma}_{\bm \Psi_k} =  \E \{\bm \Psi_k(\bm \theta_k, \bm O)~ \bm \Psi_k (\bm \theta_k, \bm O)^\top\}$. 

To estimate the assymptotic variance of $\sqrt{n}(\bm{\hat\theta}_k - \bm \theta_k)$, one first needs to estimate $\E \{ - (\bm \partial \bm \Psi_k/ \bm \partial \bm \theta_k)^{-1}\}$ and $\bm{\Sigma}_{\bm \Psi}$. Note that:
\begin{align*}
    \frac{\partial\bm\Psi_{\bm\beta_{jk}}}{\partial\bm\delta_k} &=0\\
    \frac{\partial\bm\Psi_{\bm\beta_{jk}}}{\partial\bm\beta_{j'k'}} &= \begin{cases}
        0 & \mathrm{if} ~ j\ne j' ~ \mathrm{or} ~k\ne k'\\
        I(S=k)\bm \phi(\bm L)\exp(\bm L^\top\bm \beta_{jk})\bm L & \mathrm{if} ~ j=j' ~ \mathrm{and} ~k= k'
    \end{cases}
\end{align*}
Hence, $\E \{ - (\bm \partial \bm \Psi_k/ \bm \partial \bm \theta_k)^{-1}\}$ can be easily estimated by its sample mean, using the individual-level data from trial $S=k$.

We now check whether the covariance matrix of $\bm \Psi_k$, i.e. $\bm{\Sigma}_{\bm \Psi_k} =  \Esp[\bm \Psi(\bm \theta_k, \bm O)~ \bm \Psi (\bm \theta_k, \bm O)^\top]$ can be estimated from the individual-level data of trial $k$. For this, note that:
\[\bm{\Sigma}_{\bm \Psi_k} = \E \bigg[\begin{pmatrix}
    \bm \Psi_{\beta_k} \bm \Psi_{\beta_k}^\top & \bm \Psi_{\beta_k} \bm \Psi_{\delta_k}^\top\\
    \bm \Psi_{\beta_k} \bm \Psi_{\delta_k}^\top
    & \bm \Psi_{\delta_k} \bm \Psi_{\delta_k}^\top
\end{pmatrix} \bigg]\]
The first block, $\E\{\bm \Psi_{\beta_k} \bm \Psi_{\beta_k}^\top\}$, consists of elements of the form:
\begin{align*}
&\Esp(\bm\Psi_{\bm\beta_{j,k}}\bm\Psi_{\bm\beta_{j',k}}^\top)\\ &= 
\Esp\bigg\{ \bigg[I(S=k)\bm\phi(\bm L) \exp(\bm L^T\bm\beta_{jk}) - \bm\phi(\bm L)  I(S=j)\bigg]\bigg[I(S=k)\bm \phi(L)^\top \exp(\bm L^T\bm\beta_{j'k}) - \bm\phi(\bm L)^\top I(S=j')\bigg]\bigg\}\\
&=\E\bigg\{I(S=k)\bm \phi(\bm L)\bm \phi(\bm L)^\top\exp(\bm L^T\bm\beta_{jk})\exp(\bm L^T\bm\beta_{j'k})\bigg\} -\E\bigg\{I(S=k)I(S=j')\bm \phi(\bm L)\bm \phi(\bm L)^\top\exp(\bm L^T\bm\beta_{jk})\bigg\}\\
&- \E\bigg\{I(S=k)I(S=j)\bm \phi(\bm L)\bm \phi(\bm L)^\top\exp(\bm L^T\bm\beta_{j'k})\bigg\} + \E\bigg\{I(S=j)I(S=j')\bm \phi(\bm L)\bm \phi(\bm L)^\top\bigg\}
\end{align*}
The sample analogue of the first expectation in the above expression is easily computed from the IPD of trial $S=k$. The second (and third) expectation is either $0$ when $j'\ne k$ (and $j\ne k$), or can be easily estimated by its sample analogue, using the IPD from trial $S=k$. The main challenge is in the forth expectation, which requires information on $\E\{\phi(\bm L)\bm \phi(\bm L)^\top\mid S=j\}$ for $j=1,\ldots,z$.

The second block, $\E\{\bm \Psi_{\beta_k} \bm \Psi_{\delta_k}^\top\}$, consists of elements of the form:
\begin{align*}
&\Esp\{\bm\Psi_{\bm\beta_{j,k}}\Psi_{\theta(j',k)}\}\\ &= 
\Esp\bigg\{ \bigg[I(S=k)\bm\phi(\bm L) \exp(\bm L^T\bm\beta_{jk}) - \bm\phi(\bm L)  I(S=j)\bigg]\bigg[I(S=k)~ Y [(r_{1k}^{-1} + r_{0k}^{-1})X-r_{0k}^{-1}] ~\exp(\bm \beta_{j'k}^\top \bm L) - \theta(j',k)\bigg]\bigg\}\\
&=\Esp\bigg\{ I(S=k)\bm\phi(\bm L) Y [(r_{1k}^{-1} + r_{0k}^{-1})X - r_{0k}^{-1}] \exp(2\bm L^T\bm\beta_{jk})\bigg\}\\
&-\E\bigg\{I(S=k)\bm\phi(\bm L) \exp(\bm L^T\bm\beta_{jk})\theta(j',k)\bigg\}\\
&-\E\bigg\{I(S=k) I(S=j) Y [(r_{1k}^{-1} + r_{0k}^{-1})X-r_{0k}^{-1}] ~\exp(\bm \beta_{j'k}^\top \bm L)\bm\phi(\bm L)\bigg\} \\
&+\E\bigg\{\theta(j',k)\bm\phi(L)I(S=j)\bigg\}
\end{align*}
The sample analogues of the first and second expectations in the above expression are easily computed from the IPD of trial $S=k$. The third expectation is either $0$ when $j\ne k$, or can be easily estimated by its sample analogue (using the IPD of trial $S=k$). The forth expectation can be estimated by the aggregated data of trial $S=j$.

Finally, the third block, $\E\{\bm \Psi_{\delta_k} \bm \Psi_{\delta_k}^\top\}$, consists of elements of the form:
\begin{align*}
&\Esp\{\bm\Psi_{\bm\theta({j,k})}\Psi_{\theta(j',k)}\}\\ &= 
\Esp\bigg\{ 
\bigg[I(S=k)~ Y [(r_{1k}^{-1} + r_{0k}^{-1})X-r_{0k}^{-1}] ~\exp(\bm \beta_{jk}^\top \bm L) - \theta(j,k)\bigg]\\
&\times
\E\bigg[I(S=k)~ Y [(r_{1k}^{-1} + r_{0k}^{-1})X-r_{0k}^{-1}] ~\exp(\bm \beta_{j'k}^\top \bm L) - \theta(j',k)
\bigg]\bigg\}
\end{align*}
which can be easily estimated by the corresponding sample analogues, using the IPD of trial $S=k$.
\subsection{Simulation study (section 3): residual covariance matrix}
In the simulation study of section 3, to generate the residual covariance matrix $\bm\Sigma$ (dimension $q^2 \times q^2$, where $q$ is the number of studies in the meta-analysis), we first generate an orthogonal $q^2 \times q^2$ matrix $\bm P$ by taking the QR-decomposition of a matrix whose components are randomly sampled from $\mathcal{N}(0.5,0.2)$. The $q^2$ non-negative numbers $\sigma_1, \sigma_2, \ldots, \sigma_{q^2}$ are then generated by using the sequence $\{a_n\}$: $a_1=0, a_{n+1}=a_n+5\times 10^{-3}$, $n=1$ to $q^2$. The covariance matrix $\bm\Sigma$ is then generated as:
\begin{displaymath}
\bm\Sigma = \bm P^\top \mathrm{diag}(\sigma_{q^2}, \sigma_{q^2-1}, \ldots, \sigma_1) \bm P.
\end{displaymath}
To take into account the independence between $\theta(j,k)$ and $\theta(j',k')$ when $j \neq j'$ and $k \neq k'$, the components of $\bm\Sigma$ corresponding to $\mathrm{cov}\{\theta(j,k),\theta(j',k')\}$ are switched to zero whenever $j \neq j'$ and $k \neq k'$. The whole process is conducted in \texttt{R} with the seed set at 1. $\bm\Sigma$ is then rescaled by dividing its elements by a factor of 9.9. This assures that the average variance of the random errors is approximately 0.1, hence the ratio of between- vs. within-trial is sufficiently large ($\sim 10$). The covariance matrix $\bm\Sigma$ is kept unchanged across all simulations.

\subsection{Illustrative data analysis: Patient characteristics across five trials}

        \begin{sidewaystable}
        \centering
        \caption{Data analysis: baseline characteristics}
            \scalebox{0.7}{\begin{tabular}{lrrrrrrrrrr}
            \hline
            Characteristics & \multicolumn{2}{c}{Study 1} & \multicolumn{2}{c}{Study 2} & \multicolumn{2}{c}{Study 3} & \multicolumn{2}{c}{Study 4} & \multicolumn{2}{c}{Study 5} \\
            \hline
            Treatment & Risankizumab & Ustekinumab & Risankizumab & Ustekinumab & Risankizumab & Ustekinumab & Risankizumab & Ustekinumab & Risankizumab & Ustekinumab \\
            \quad No. (\%) & 52 (72.2)& 20 (27.8)& 91 (78.4) & 25 (21.6) & 118 (73.3) & 43 (26.7) & 85 (75.9) & 27 (24.1) & 133 (76.0) & 42 (24.0)\\
            \hline
            Covariates & & & & & & & & & & \\
            \quad Men - no. (\%) & 41 (78.9) & 17 (85.0) & 68 (74.7) & 18 (72.0) & 73 (61.9) & 26 (60.5) & 58 (68.2) & 18 (66.7) & 96 (72.2) & 28 (66.7) \\
            \quad Age - Mean (SD) & 46.3 (12.0) & 49.1 (12.8) & 46.7 (13.3) & 43.8 (14.3) & 50.4 (14.2) & 49.1 (12.8) & 47.3 (13.4) & 48.9 (15.2) & 47.3 (13.4) & 53.1 (14.4) \\
            \quad BMI - Mean (SD) & 27.9 (5.8) & 28.3 (6.3) & 29.6 (7.0) & 30.3 (7.1) & 31.2 (7.3) & 31.1 (7.2) & 31.7 (7.4) & 31.8 (7.0) & 31.7 (6.8) & 33.1 (6.9) \\
            \quad Baseline PASI - Mean (SD) & 23.0 (9.7) & 21.3 (7.5) & 18.7 (5.8) & 18.7 (7.0) & 21.3 (7.9) & 20.4 (6.8) & 20.0 (7.2) & 18.6 (5.1) & 20.3 (8.9) & 18.3 (6.5) \\
            \hline
            Outcome & & & & & & & & & & \\
            \quad PASI week 12 – Mean (SD) & 1.6 (2.2) & 3.2 (2.5) & 1.9 (2.6) & 2.6 (2.5) & 2.9 (5.0) & 4.8 (6.3) & 2.0 (2.7) & 3.3 (3.9) & 2.3 (3.5) & 3.4 (4.1) \\
            \quad PASI 75 – no. (\%) & 49 (94.2) & 15 (75.0) & 82 (90.1) & 21 (84.0) & 100 (84.8) & 25 (58.1) & 80 (94.1) & 19 (70.4) & 116 (87.2) & 29 (69.1) \\
            \hline
            \end{tabular}
        \label{tab:4}
        }
        \end{sidewaystable}

\end{document}